Superconducting phase diagram of Sb and Se substitution in CeOBiS$_2$ single crystals

Tatsuya Suzuki[1], Masanori Nagao[1*], Yuki Maruyama[1], Satoshi Watauchi[1], and Isao Tanaka[1]

[1]*University of Yamanashi, 7-32 Miyamae, Kofu, Yamanashi 400-0021, Japan*

*Corresponding Author

Masanori Nagao

Postal address: University of Yamanashi, Center for Crystal Science and Technology

Miyamae 7-32, Kofu, Yamanashi 400-0021, Japan

Telephone number: (+81)55-220-8610

Fax number: (+81)55-220-8270

E-mail address: mnagao@yamanashi.ac.jp

**Abstract**

Sb- and Se-substituted CeOBiS$_2$ single crystals have been successfully grown using CsCl/KCl flux. Sb and Se substitution dependence of the superconductivity on CeOBiS$_2$ was investigated and the superconducting phase diagram at above 0.3 K was described. The non-linear boundary between superconductivity and non-superconductivity was revealed.

**Main text**

**1. Introduction**

Layered superconductors exhibited high transition temperatures such as cuprate[1] and iron-based superconductors.[2] $CeOBiS_2$ is one of the $BiS_2$-based layered superconductors such as $La(O,F)BiS_2$ [3] which is a layered superconductor. Several $BiS_2$-based layered compounds with superconductivity require the F substitution in the O-site for the electron carrier introduction. Among them, $CeOBiS_2$ without F substitution exhibits superconductivity by the Ce valence fluctuation.[4,5] The Ce valence fluctuation formed at of $Ce^{3+}$ and $Ce^{4+}$ mixing valence state was observed by X-ray photoelectron spectroscopy (XPS)[5] and X-ray absorption spectroscopy (XAS).[6] Furthermore, the temperature dependence of resistivity at a normal state was reported weak semiconducting-like behavior.[4,5] On the other hand, the $CeOBiS_2$ with Arrhenius-type behavior at a normal state resistivity was reported no superconductivity.[7] This difference was explained by the existence of two kinds of local structure configurations in the $CeOBiS_2$ which intrinsically exhibited semiconducting behavior, but a metallic phase appeared due to the local distortions.[8] Therefore, we focused on the superconducting $CeOBiS_2$ which was a simple chemical formulation in $BiS_2$-based layered superconductors. Previously we found the superconductivity was suppressed with Sb substitution into the Bi-site, which is presumed to originate from the change of the crystal system.[9] In contrast, the CeOBiS$_2$ superconductor with Se substitution into the S-site exhibited enhancement of in-plane chemical pressure[10], suppression of in-plane disorder, and decrease of Ce ions valence. As a result, an increase in superconducting transition temperature was observed.[11] Especially, the in-plane chemical pressure increases the density of states at the Fermi level, which may increase the phonon frequency. The possibility of the enhancement of superconductivity by these phenomena was suggested.[10]

In this paper, the effect of both Sb and Se substitutions for superconductivity at above 0.3 K was systematically investigated using the Sb- and Se-substituted CeOBiS$_2$ [CeO(Bi,Sb)(S,Se)$_2$] single crystals. Then superconducting phase diagram of Sb and Se substitution in CeOBiS$_2$ was revealed.

## 2. Experimental Details

Sb- and Se-substituted CeOBiS$_2$ [CeO(Bi,Sb)(S,Se)$_2$] single crystals were grown using CsCl/KCl flux.[4,12,13] The raw materials of Ce$_2$S$_3$, Bi$_2$O$_3$, Bi$_2$S$_3$, Bi, Sb (or Sb$_2$O$_3$), and Se were weighed to a total amount of 0.8 g for a nominal composition of CeOBi$_{1-x}$Sb$_x$S$_{2-y}$Se$_y$ ($0 \leq x \leq 0.20$, $0.125 \leq y \leq 0.500$). The molar ratio of the CsCl/KCl flux was CsCl:KCl = 5:3 with a total amount of 5.0 g. The raw materials (0.8 g) and CsCl/KCl flux (5.0 g) were mixed and ground using a mortar and then sealed in an evacuated quartz tube (~10 Pa). The prepared quartz tube was heated at 950 °C (Except for $x = 0$, $y = 0.500$, which was performed at 1000 °C.) for 10 h, followed by cooling to 600 °C at a rate of 1 °C/h, then the sample was spontaneously cooled down to room temperature (~30 °C) in the furnace. The heated quartz tube was opened to air and the obtained materials were washed and filtered to remove the CsCl/KCl flux using distilled water.

The obtained crystals confirmed the CeOBiS$_2$ structure[5)] by X-ray diffraction (XRD) (Rigaku; MultiFlex) with CuK$\alpha$ radiation. The compositional ratio of the grown crystals was evaluated using energy dispersive X-ray spectrometry (EDS) (Bruker; Quantax 70) associated with the observation of the microstructure, based on scanning electron microscopy (SEM) (Hitachi High-Technologies; TM3030). Analytical compositions of each element were defined as $C_{XX}$ (XX: Bi, Sb, S, and Se). The obtained compositional values were normalized using $C_S + C_{Se} = 2$ (S + Se analytical composition was 2), and then Bi, and Sb compositions ($C_{Bi}$, and $C_{Sb}$) were determined.

The picked single crystals at random from each sample lot were measured the resistivity−temperature ($\rho$−$T$) characteristics. The $\rho$−$T$ characteristics of the picked single crystals were determined using the standard four-probe method with a constant current density mode range of 20−50 mA/cm$^2$ using a physical property measurement system (Quantum Design; PPMS DynaCool). The electrical terminals were fabricated using Ag paste (DuPont; 4922N). The $\rho$–$T$ characteristics in the temperature range of 0.3−15 K were evaluated based using the adiabatic demagnetization refrigerator (ADR) option for the PPMS. A magnetic field of 3.0 T at 1.9 K was applied to operate the ADR, which was subsequently removed. Consequently, the temperature of the sample decreased to approximately 0.3 K. The measurement of $\rho$−$T$ characteristics was initiated at the lowest temperature (~0.3 K), which was spontaneously increased to 15 K. The superconducting transition temperature with zero resistivity ($T_c^{zero}$) was estimated from the $\rho$–$T$ characteristics. The $T_c^{zero}$ was determined as the temperature at which the resistivity is below approximately 300 μΩcm (Except for $C_{Sb}$ = 0.068, $C_{Se}$ = 0.28, which was determined at 3.0 mΩcm due to a technical problem.). The compositional ratio of the $\rho$−$T$ characteristics measured samples had been evaluated by EDS, and then they were employed for the superconducting phase diagram.

## 3. Results and Discussion

The obtained CeO(Bi,Sb)(S,Se)$_2$ single crystals exhibited a plate-like shape with a size of approximately 1.0 mm and a thickness of 100−200 µm, which were similar in shape to Sb-substituted CeOBiS$_2$ [CeO(Bi,Sb)S$_2$] single crystals.[9] However, the well-developed plane in CeO(Bi,Sb)(S,Se)$_2$ single crystals was rougher than that of CeO(Bi,Sb)S$_2$ single crystals. A well-developed plane in the obtained CeO(Bi,Sb)(S,Se)$_2$ single crystals corresponded to the *c*-plane of CeOBiS$_2$ structure[4] by the XRD patterns. The *c*-axis lattice parameters for the obtained CeO(Bi,Sb)(S,Se)$_2$ single crystals were range of 13.53−13.62 Å with depending on both Sb and Se substitution amount. Analytical compositions of Sb and Se in the obtained CeO(Bi,Sb)(S,Se)$_2$ single crystals were lower than those of the nominal compositions, and these analytical compositions exhibited dispersion in the same grown lot. Table I summarized the nominal (*x* and *y*) and analytical ($C_{Bi}$, $C_{Sb}$, $C_S$, and $C_{Se}$) compositions with normalized using $C_S + C_{Se} = 2$ for the CeO(Bi,Sb)(S,Se)$_2$ single crystals. All samples exhibited the Bi-site deficiency which was observed $C_{Bi} + C_{Sb} < 1$. The amount of Bi-site deficiency was in the range of 4−9 at% which was a similar range of the only Sb-substituted CeOBiS$_2$ [CeO(Bi,Sb)S$_2$] single crystals.[9] The Bi-site deficiency values were higher than CeOBiS$_2$ without Sb substitution,[14,15] this indicated that Sb substitution enhanced the Bi-site deficiency in the CeOBiS$_2$ structure. Besides, we investigated the compositions of the obtained CeO(Bi,Sb)(S,Se)$_2$ single crystals into the difference of Sb source which was Sb or $Sb_2O_3$. And then this result hardly changed.

The CeO(Bi,Sb)(S,Se)$_2$ single crystals at random in each lot grown from each nominal composition were picked, and the $\rho$–$T$ measurements and compositional analysis were performed for the superconducting phase diagram investigation. The $\rho$−$T$ characteristics of some typical CeO(Bi,Sb)(S,Se)$_2$ single crystals were shown in Figure 1. Weak semiconducting-like behavior at near above superconducting transition temperature in a normal state was observed. At first glance, a correlation between normal state resistivity and superconducting transition temperature seems to exist. However, in the case of the samples without superconducting transition (*Ex.* $C_{Sb}$ = 0.05, $C_{Se}$ = 0.09), the normal state resistivity was lower than that of superconducting samples. Then the relationship between normal state resistivity and superconductivity was unclear in this experiment. Figure 2 shows the superconducting phase diagram of Sb and Se substitution dependence for CeO(Bi,Sb)(S,Se)$_2$ single crystals. The data of CeO(Bi,Sb)S$_2$ single crystals were referred from Ref. 9. The superconducting transition temperature with zero resistivity ($T_c^{zero}$) was decreased with the increase in Sb substitution and increased with the increase in Se substitution. Superconductivity in CeOBiS$_2$ was suppressed as the Sb substitution, enhanced by the Se substitution. Those behaviors exhibited the same trend as previous reports.[9,11] In other words, the Sb substitution region with superconducting observation was spread by the Se substitution. The boundary between superconductivity and non-superconductivity at above 0.3 K showed non-linear which is plotted in the dashed line in Figure 2 (b). That boundary showed the dome shape at $C_{Se}$ = 0.15−0.30 region which increased the $T_c^{zero}$. Se-substituted CeOBiS$_2$ [CeOBi(S,Se)$_2$] superconducting phase diagram showed a similar dome shape which originated from the in-plane chemical pressure effects[10] and decrease of carrier concentration due to the Ce valence state.[11] We assumed that the dome shape behavior of superconducting transition temperature in Figure 2 (b) originated from the same phenomena.

## 4. Conclusion

Sb- and Se-substituted CeOBiS$_2$ [CeO(Bi,Sb)(S,Se)$_2$] single crystals were grown and measured the superconducting transition temperature. A superconducting phase diagram with $C_{Sb}$ < 0.18, $C_{Se}$ < 0.45 for CeOBiS$_2$ was revealed. The boundary between superconductivity and non-superconductivity at above 0.3 K showed a non-linear line with the dome shape at $C_{Se}$ = 0.15−0.30 region.

**Acknowledgments**

This work was supported by JSPS KAKENHI (Grant-in-Aid for Scientific Research (B) and (C): Grant Number 21H02022, 19K05248, and 23K03358, Grant-in-Aid for Challenging Exploratory Research: Grant Number 21K18834).

**Figure caption**

Figure 1 (Color online) Resistivity−temperature ($\rho-T$) characteristics of typical CeO(Bi,Sb)(S,Se)$_2$ single crystals.

Figure 2 (Color online) (a) Superconducting phase diagram of Sb- and Se-substituted composition ($C_{Sb}$ and $C_{Se}$) for CeOBiS$_2$ single crystals (b) projection of $C_{Sb}-C_{Se}$ plane. The data of CeO(Bi,Sb)S$_2$ single crystals were referred from Ref. 9.

Table I  Nominal Sb and Se composition ($x$ and $y$), and analytical composition ($C_{Bi}$, $C_{Sb}$, $C_S$, and $C_{Se}$) in the obtained crystals. ($C_{Bi}$, $C_{Sb}$, $C_S$, and $C_{Se}$ compositions were normalized by $C_S + C_{Se} = 2$.)

| Nominal composition | | Analytical composition (Normalized using $C_{Sb} + C_{Se} = 2$) | | | |
|---|---|---|---|---|---|
| Sb:$x$ | Se:$y$ | $C_{Bi}$ | $C_{Sb}$ | $C_S$ | $C_{Se}$ |
| 0 | 0.125 | 0.95±0.02 | 0.00±0.00 | 1.89±0.02 | 0.11±0.02 |
| 0.050 | 0.125 | 0.92±0.02 | 0.02±0.01 | 1.91±0.02 | 0.09±0.02 |
| 0.050* | 0.125 | 0.93±0.04 | 0.03±0.02 | 1.92±0.03 | 0.08±0.03 |
| 0.075 | 0.125 | 0.89±0.04 | 0.05±0.01 | 1.91±0.02 | 0.09±0.02 |
| 0.100 | 0.125 | 0.85±0.01 | 0.08±0.03 | 1.87±0.01 | 0.13±0.01 |
| 0.100* | 0.125 | 0.87±0.02 | 0.05±0.02 | 1.91±0.02 | 0.09±0.02 |
| 0.150* | 0.125 | 0.82±0.02 | 0.11±0.02 | 1.90±0.01 | 0.10±0.01 |
| 0 | 0.250 | 0.96±0.02 | 0.00±0.00 | 1.79±0.02 | 0.21±0.02 |
| 0.050 | 0.250 | 0.90±0.03 | 0.02±0.01 | 1.80±0.02 | 0.20±0.02 |
| 0.075 | 0.250 | 0.88±0.04 | 0.05±0.01 | 1.81±0.01 | 0.19±0.01 |
| 0.100 | 0.250 | 0.87±0.04 | 0.09±0.05 | 1.84±0.02 | 0.16±0.02 |
| 0.150 | 0.250 | 0.84±0.06 | 0.11±0.05 | 1.83±0.06 | 0.17±0.06 |
| 0 | 0.375 | 0.95±0.01 | 0.00±0.00 | 1.70±0.01 | 0.31±0.01 |
| 0.050 | 0.375 | 0.93±0.02 | 0.02±0.01 | 1.72±0.02 | 0.29±0.02 |
| 0.075 | 0.375 | 0.86±0.04 | 0.06±0.04 | 1.63±0.05 | 0.37±0.05 |
| 0.100 | 0.375 | 0.89±0.02 | 0.04±0.01 | 1.70±0.02 | 0.29±0.02 |
| 0.125 | 0.375 | 0.82±0.09 | 0.09±0.05 | 1.72±0.04 | 0.28±0.04 |
| 0.150 | 0.375 | 0.83±0.05 | 0.11±0.06 | 1.66±0.02 | 0.34±0.02 |
| 0 | 0.500 | 0.94±0.01 | 0.00±0.00 | 1.59±0.03 | 0.41±0.03 |
| 0.050 | 0.500 | 0.90±0.02 | 0.04±0.02 | 1.63±0.02 | 0.37±0.02 |
| 0.100 | 0.500 | 0.88±0.04 | 0.06±0.01 | 1.60±0.02 | 0.40±0.02 |
| 0.125 | 0.500 | 0.86±0.02 | 0.08±0.01 | 1.59±0.03 | 0.41±0.03 |
| 0.150 | 0.500 | 0.84±0.06 | 0.11±0.03 | 1.60±0.03 | 0.40±0.03 |
| 0.200 | 0.500 | 0.77±0.04 | 0.15±0.02 | 1.57±0.03 | 0.43±0.03 |

*: Sb source was employed as $Sb_2O_3$.

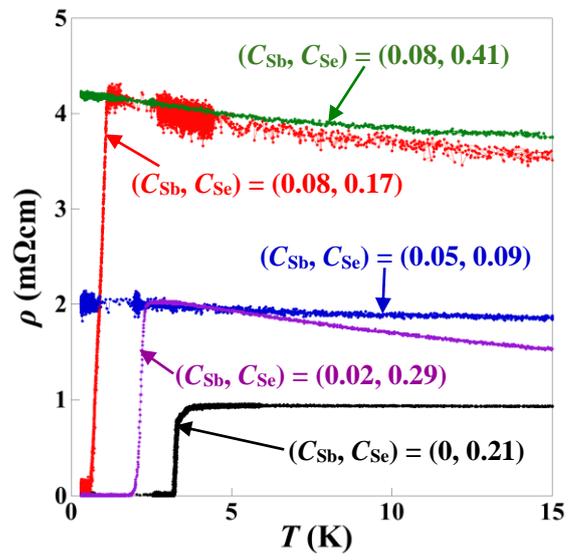

**Figure 1**

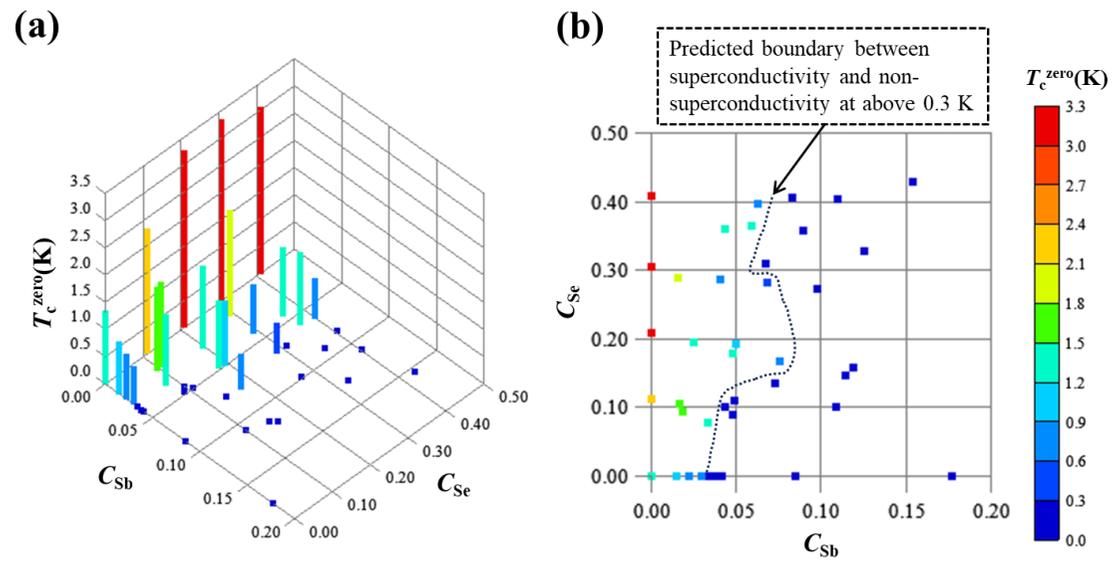

**Figure 2**